%% file: main.tex
\documentclass[sigplan, nonacm]{acmart}

\usepackage{tikz}
\usepackage{amsmath}
\usepackage{filecontents}
\usepackage{ifthen}
\usepackage{booktabs}
\usepackage{subcaption}
\usepackage{enumitem}
\usepackage{ulem}
\usepackage{dblfloatfix}
\usepackage{graphicx}
\usepackage{multirow}
\usepackage{comment}
\usepackage{booktabs}
\usepackage{pifont}
\usepackage{graphicx}
\usepackage{xcolor}
\newboolean{publicversion}
\setboolean{publicversion}{true}

\ifthenelse{\boolean{publicversion}}{
 \newcommand{\grumbler}[3]{}
 \newcommand{\jm}[1]{}
 \newcommand{\alind}[1]{}
}
{
\newcommand{\grumbler}[3]{\xspace\textcolor{#3}{\bf #1: #2}}
\newcommand{\jm}[1]{\grumbler{Jayashree}{#1}{red}}
\newcommand{\alind}[1]{\grumbler{Alind}{#1}{green}}

}
\newcommand{\cmark}{{\color{green!60!black}\ding{51}}}
\newcommand{\lmark}{{\color{orange!}\ding{51}}}   
\newcommand{\xmark}{{\color{red!80!black}\ding{55}}}
\usepackage{xspace}

\input{macros}

\begin{document}
\setlength{\textfloatsep}{8pt plus 2pt minus 4pt}
\setlength{\floatsep}{8pt plus 2pt minus 4pt}
\setlength{\intextsep}{8pt plus 2pt minus 4pt}
\setlength{\dbltextfloatsep}{8pt plus 2pt minus 4pt}
\setlength{\dblfloatsep}{8pt plus 2pt minus 4pt}

\title[]{Sutradhara: An Intelligent Orchestrator-Engine Co-design for Tool-based Agentic Inference}

\author{Anish Biswas}
\affiliation{%
  \institution{Microsoft Research India}
  \city{Bengaluru}
  \country{India}
}

\author{Kanishk Goel}
\affiliation{%
  \institution{Microsoft Research India}
  \city{Bengaluru}
  \country{India}
}

\author{Srivarshinee S}
\affiliation{%
  \institution{Microsoft M365 Research}
  \city{Bengaluru}
  \country{India}
}

\author{Jayashree Mohan}
\affiliation{%
  \institution{Microsoft Research India}
  \city{Bengaluru}
  \country{India}
}

\author{Alind Khare}
\affiliation{%
  \institution{Microsoft M365 Research}
  \city{Bengaluru}
  \country{India}
}

\author{Anjaly Parayil}
\affiliation{%
  \institution{Microsoft M365 Research}
  \city{Bengaluru}
  \country{India}
}

\author{Chetan Bansal}
\affiliation{%
  \institution{Microsoft M365 Research}
  \city {Redmond}
  \country{USA}
}

\author{Ram Ramjee}
\affiliation{%
  \institution{Microsoft Research India}
  \city {Redmond}
  \country{USA}
}
\renewcommand{\shortauthors}{}

\input{sections/abstract}
\maketitle
\input{sections/intro}

\input{sections/bgk}
\input{sections/analysis}
\input{sections/design}
\input{sections/eval}
\input{sections/related}
\input{sections/conclusion}

\bibliographystyle{ACM-Reference-Format}
\bibliography{bibs/related_work}

\appendix
\input{sections/appendix}

\end{document}

%% file: macros.tex
\newcommand{\sysname}{\textsc{Sutradhara}\xspace}
\newcommand{\sysnameshort}{\textsc{SD}\xspace}

\newcommand{\sref}[1]{(\S\ref{#1})}

\newcommand{\jheading}[1]{\vspace{0.05in}\noindent\textbf{#1.}}
\newcommand{\hlab}[1]{}
\newcommand{\jheadingita}[1]{\vspace{0.05in}\noindent\uline{\textit{#1.}}}

\newenvironment{tightitemize}{%
\begin{list}{$\bullet$}{%
\setlength{\itemsep}{1.5pt}%
\setlength{\topsep}{2pt}%
\setlength{\parskip}{0pt}%
\setlength{\parsep}{0pt}%

\setlength{\labelwidth}{0pt}%
\setlength{\leftmargin}{4pt}%
\setlength{\labelsep}{0pt}%
\setlength{\listparindent}{0pt}%
}}%
{\end{list}}

{\end{list}}

%% file: sections/abstract.tex
\begin{abstract}
Agentic applications are LLMs  that iteratively invoke external tools to accomplish complex tasks \cite{react, wang2023survey}. Such tool-based agents are rapidly becoming the dominant paradigm for deploying language models in production. Unlike traditional single-turn inference, agentic workloads chain together multiple LLM calls and tool executions before producing a final response, creating a new performance bottleneck that manifests as increased latency in First Token Rendered (FTR) of the final answer. Through analysis of requests at production scale, we reveal three critical challenges: tool calls account for 30-85\% of FTR latency, KV cache hit rates collapse despite substantial context reuse across iterations \cite{zheng2023sglang, vllm_sosp}, and sequential orchestration wastes potential intra-request parallelism. 
These bottlenecks stem from a design gap in which orchestrators and LLM engines operate as decoupled black boxes, preventing cross-layer optimizations.

We present \sysname, a co-designed agentic inference system that integrates orchestration with LLM serving through a thin API enabling three optimizations: overlap tool execution with subsequent LLM prefill using tool-aware prompt splitting, streaming tool execution to dispatch tools incrementally during decode rather than waiting for complete output, and orchestrator-aware cache management that uses semantic hints to improve hit rates and reduce thrashing. Implemented on vLLM, 
\sysname{} improves the throughput–latency trade-off in agentic systems, sustains up to 77\% higher load at the same median FTR latency, or reduces median FTR latency by up to 15\% at the same load while reducing end-to-end latency by up-to 11\% on A100 GPUs.
\end{abstract}




%% file: sections/intro.tex
\section{Introduction}
Large language models have rapidly evolved from research prototypes to production systems powering mission-critical applications. While early deployments focused on simple query-response patterns, modern LLM applications increasingly adopt agentic architectures—autonomous systems that iteratively invoke LLMs and external tools to accomplish complex tasks. These systems represent a fundamental shift in how we deploy LLMs. Rather than treating models as stateless oracles that answer isolated queries, agentic architectures enable LLMs to reason, plan, and interact with the external world through iterative tool use \cite{react, miao2024llm}.

\input{fig-tex/intro-fig}

However, this architectural shift introduces a critical performance challenge that existing LLM serving infrastructure fails to address: latency explosions in user-perceived response times. Traditional LLM serving systems optimize for Time-To-First-Token (TTFT) and inter-token latency, treating each inference request as an independent unit of work. In contrast, agentic applications chain together multiple LLM calls and tool invocations in iterative loops before producing a final response. User-perceived latency—measured as First Token Rendered (FTR) of the final answer encompasses not just a single LLM forward pass, but the cumulative latency of multiple LLM calls interspersed with tool executions until the first user-visible token is generated. In production systems, FTR latency can span seconds or even tens of seconds severely degrading user experience. 

To understand these challenges systematically, we conduct the first large-scale empirical study of agentic inference performance using synthetic traces from production workloads in a large cloud provider. Our analysis reveals three critical insights that challenge conventional assumptions about LLM serving bottlenecks. (1) \textit{Tool execution dominates tail latency}, accounting for 30-85\% of FTR latency while individual tool calls can exceed LLM prefill time. While LLM prefill and decode have been the traditional focus of optimization efforts~\cite{parrot, autellix, continuum, vllm_sosp, sarathi}, we find that tool calls—often dismissed as "thin" external I/O operations—could be a major component of FTR latency. (2) \textit{Sequential orchestration leaves parallelism unexploited}, as 60-80\% of each iteration's prefill is tool-independent yet current systems wait for all tool outputs before beginning the next iteration; similarly tools are executed only after LLM decodes complete, while it is possible to stream tool executions as decode progresses. (3) \textit{KV cache thrashing destroys reuse opportunities} 
despite context reuse, 
as workload-agnostic LRU eviction thrashes shared prefixes when agentic requests execute concurrently.

These findings point to a fundamental architectural mismatch: \textbf{orchestrators and LLM engines operate as decoupled black boxes}, communicate only through opaque request-response interfaces. The orchestrator has knowledge of iterations, tool dependencies, and prompt composition, while the engine controls scheduling, batching, and KV cache management. Neither layer leverages information from the other to make globally optimal decisions. Breaking this abstraction barrier is essential to address 
bottlenecks.

We present \sysname, a co-designed agentic inference system that tightly integrates orchestration logic with LLM engine scheduling through a thin, principled API. 
\sysname enables three key optimizations: (1) parallel execution through prompt splitting, where iteration $i+1$ speculatively begins prefill using tool-independent context while iteration $i$'s tools execute, then incrementally extends the prefill when tool outputs arrive; (2)
streaming tool dispatch during decode, where a streaming JSON parser identifies complete tool call objects as they are decoded and dispatches them immediately rather than waiting for full decode completion; and (3) orchestrator-aware cache management, where semantic metadata tags and reuse hints guide a priority-aware eviction policy that retains high-value blocks while evicting transient content, combined with request-aware scheduling that prioritizes completing in-flight requests. Figure~\ref{fig:intro} shows how intra request parallelism enables \sysname to achieve significant latency reduction of up-to 42\% per request.

We implement \sysname as an asynchronous event-driven orchestrator built on top of vLLM v0.11.0, with targeted modifications to the v1 scheduler totaling fewer than 3500 lines of code. The orchestrator-engine interface consists of five new API calls (Table~\ref{table:codesign_api}) that enable hint passing, parallel execution coordination, and cache metadata communication. Our implementation requires no changes to model architectures, training procedures, or inference kernels. We evaluate \sysname across A100-80GB GPUs using synthetic traces from production on  Qwen3-14B. Compared to the vLLM baseline, 
\sysname{} achieves a better throughput–latency tradeoff, sustains up to 77\% higher load at the same median FTR latency, or reduces median FTR latency by up to 15\% at the same load while improving end-to-end latency by 11\%.

\noindent In summary, this paper makes the following contributions:
\begin{tightitemize}
    \item \textbf{Empirical characterization.} First large-scale analysis of agentic inference workloads, identifying tool execution variance, sequential bottlenecks, and cache thrashing as dominant sources of tail FTR latency \sref{sec:analysis}.
    \item \textbf{Co-designed architecture.} We propose \sysname, an orchestrator-engine co-design, with three novel techniques; prompt splitting with tool execution overlap, streaming tool dispatch, and semantic cache management that exploits agentic request structure \sref{sec:design}.
    \item \textbf{Implementation and evaluation.} 
    We implement \sysname{} on vLLM with minimal invasiveness  and demonstrate across models and workloads that it sustains up to 77\% higher load at the same median FTR latency, or alternatively reduces median and tail FTR latency by up to 15\% and 11\% at the same load on an A100-GPU cluster~\sref{sec:eval}.
\end{tightitemize}

%% file: fig-tex/intro-fig.tex



\begin{figure}[t]
    \centering
    \begin{subfigure}[b]{0.95\linewidth}
        \includegraphics[width=\linewidth]{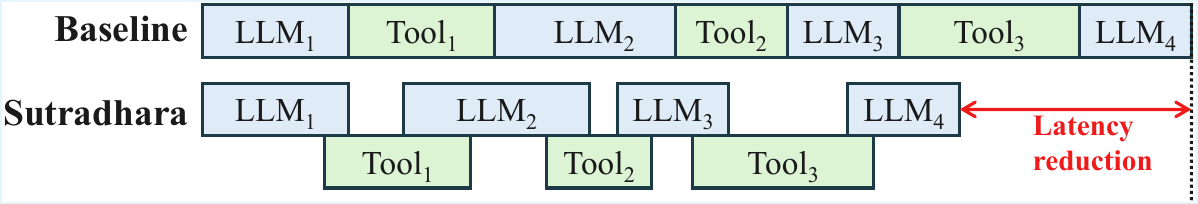}
        \caption{\small \sysname systematically unlocks intra-request parallelism}
    \end{subfigure}
    \vspace{-0.15cm}
    \begin{subfigure}[b]{0.24\linewidth}
        \includegraphics[width=\linewidth]{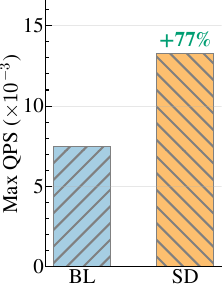}
        \caption{Throughput}
    \end{subfigure}
    \hfill
    \begin{subfigure}[b]{0.24\linewidth}
        \includegraphics[width=\linewidth]{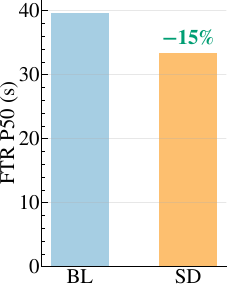}
        \caption{FTR Latency}
    \end{subfigure}
    \hfill
    \begin{subfigure}[b]{0.44\linewidth}
        \includegraphics[width=\linewidth]{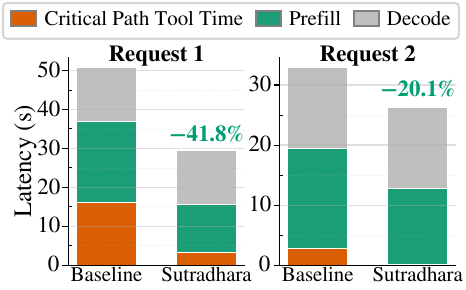}
        \caption{FTR Breakdown}
    \end{subfigure}
    \caption{\small (a)\sysname systematically parallelizes the execution of LLM and tools and enables workload-aware KV eviction.  (b) \sysname(\sysnameshort) sustains upto 77\% higher load than baseline (BL) at the same median FTR ($38s$) and (c) reduces median FTR by 15\% at the same load. (d) For two random requests in the trace, these techniques reduce FTR by 20 -- 42\%.} 
    \label{fig:intro}
\end{figure}

%% file: sections/bgk.tex
\section{Background}

\subsection{Agentic Inference}
Agentic inference represents a fundamental departure from traditional single-turn LLM inference. In standard LLM serving, a user submits a query, the model performs a single forward pass (prefill followed by autoregressive decode), and returns a complete response. The serving system optimizes for metrics like Time-To-First-Token (TTFT) and per-token decode latency, treating each request as an isolated, stateless computation.
Agentic inference, by contrast, enables LLMs to interact with external tools and APIs through iterative execution loops. Rather than producing a final answer in one shot, the LLM reasons about a task, decides which tools to invoke, examines their outputs, and continues this process until reaching a satisfactory answer.

\subsection{Agentic inference serving framework}
Figure~\ref{fig:serving_architecture} illustrates the standard architecture for agentic inference serving, consisting of three primary components that interact through well-defined interfaces.

\jheading{LLM Serving Engine} The engine handles low-level inference execution—batching requests, managing GPU memory, scheduling prefill and decode operations.
Modern production systems use frameworks like vLLM \cite{vllm_sosp} or SGLang \cite{zheng2023sglang}, which implement optimizations such as continuous batching, PagedAttention, and prefix caching. 
The engine exposes a request-response API where clients submit prompts and receive generated token sequences.
Importantly, the engine treats each LLM call as independent—it has no visibility into whether a request is part of an agentic workflow, or which iteration it belongs to.

\jheading{Orchestrator} The orchestrator implements the control logic for agentic execution. Built as an asynchronous event-driven system (typically using Python's asyncio or similar frameworks), it manages the iterative loop: constructing prompts from conversation history and tool outputs, dispatching LLM calls to the engine, parsing structured outputs to extract tool specifications, invoking tools through their respective APIs, and tracking iteration state. The orchestrator maintains semantic knowledge about request structure—which prompt sections depend on tool outputs, which tools are executing
, and how context accumulates across iterations. However, this knowledge remains isolated within the orchestrator and is not communicated to the engine.

\jheading{Tool Execution Layer} Tools are services accessed through APIs, libraries, or sand-boxed execution environments \cite{qin2024tool}. Common tools include web search, code execution, file system operations, and API calls to third-party services. Tools execute asynchronously and independently; when multiple tools are invoked in the same iteration, they run in parallel \cite{kim2024llmcompiler}. Tool execution time varies dramatically based on query complexity, external service load, and network conditions, ranging from milliseconds to seconds.

\input{fig-tex/bgk}

\jheading{Multi-iteration execution structure} An agentic request consists of multiple iterations, where each iteration $i$ follows a structured pattern:
\begin{enumerate}
    \item LLM Call: The orchestrator submits a prompt to the LLM engine containing system instructions, conversation history, and (for $i>1$) outputs from previous tool calls. The engine performs prefill to process the prompt, then generates tokens auto-regressively during decode. For the intermediate iterations, the decode output specifies which tools to invoke, typically formatted as structured JSON. For the final iteration, the decode output is the user-facing final response.
    \item Tool execution : The orchestrator parses the tool specifications and dispatches each tool for execution. Each iteration can dispatch multiple tools, that may execute independently and finish at different times. 
\end{enumerate}

\jheading{Sequential execution} Current agentic systems enforce a strict sequential execution between prefills, decodes, and tool calls per iteration. Our analysis shows that there is scope to enable intra-request parallelism across iterations, thereby optimizing request completion latencies.


%% file: fig-tex/bgk.tex
\begin{figure}[t]
    \centering
\includegraphics[width=0.7\linewidth]{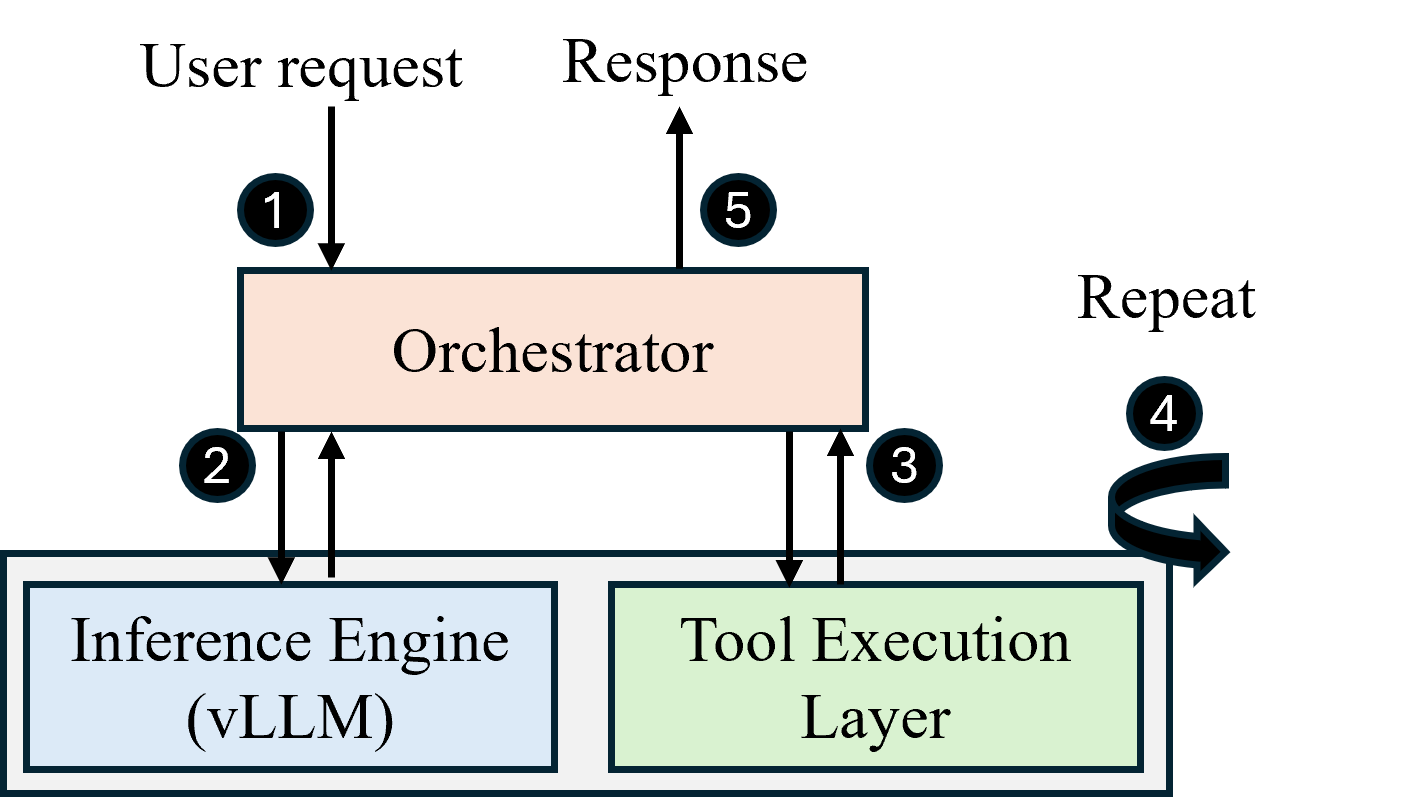}
\caption{\small Workflow: (1) User request arrives; (2) Orchestrator sends LLM query and receives response; (3) Tools optionally invoked based on response; (4) Iterative loop of LLM and tools; (5) Final user-visible response returned.}
    \label{fig:serving_architecture}
\end{figure}

%% file: sections/analysis.tex
\section{Analysis}
\label{sec:analysis}
\input{fig-tex/combined-analysis}

This section presents the agentic workload characterization that highlights bottlenecks in orchestration and KV cache behavior. We conclude with design requirements for an optimized system.

\subsection{Trace collection}
\jheading{Agentic Platform} To understand the performance characteristics of agentic inference in production environments, we analyze workloads from a major cloud provider's internal agentic platform deployed across thousands of enterprise customers. This platform provides a general-purpose orchestration framework that manages iterative LLM-tool execution loops. It dispatches requests to a heterogeneous fleet of LLM serving endpoints and coordinates calls to an extensible tool registry containing 20+ integrated services including web search APIs, enterprise chat, email, file searches, code execution, internal knowledge bases, and third-party SaaS integrations. The orchestrator implements a standard agentic execution pattern: issue an LLM request with available tools, parse structured output to identify tool calls, execute tools and collect outputs, append results to context, and iterate until the LLM produces a final response. The orchestrator could modify instructions and system prompts in any iteration depending on the set of tool calls made. This architecture is representative of popular agentic systems including LangChain~\cite{langchain}, AutoGen~\cite{autogen}, and proprietary enterprise frameworks.

\jheading{Workload Generation} We analyze workloads issued on this platform with synthetic user profiles and data, instead of the real customer-facing queries as its prompt tokens are eyes-off for privacy reasons. The synthetic queries on this actual production platform are created through an automated pipeline that first provisions test users along with their synthetically generated enterprise context, including files, chats, and calendar events. Then, a structured sets of evaluation queries are executed on the platform, each designed to replicate realistic tasks such as document retrieval, summarization, and information search. By modeling workflows and entity relationships, the synthetic workload environment closely mirrors production scenarios. 
We also validate all the observations gained from synthetic workloads against customer-facing production workload. 
We analyze 6000 agentic requests across different user categories.

\subsubsection{Metrics and Definitions}
\label{sec:metrics}
The key metrics are:

\jheadingita{First Token Rendered (FTR)}
\emph{FTR} is defined as the elapsed time from user request submission to the rendering of the \emph{first token} of the final user-visible response. 

\jheadingita{Iteration Depth} No. of LLM calls in an agentic request.

\jheadingita{Tool Call Fan-Out} Number of tool invocations \emph{per iteration}.

\subsection{Trace statistics}
We first present the overall trace characteristics. For each request, we categorize the number of LLM iterations into two types; intermediate iterations that perform tool invocations and a final iteration that generates the user visible response and doesn't result in any tool calls. Figure~\ref{fig:trace_analysis} presents the overall trace characteristics (a) the distribution of iteration depth per request (b) tool call fan out per iteration, (c) prefill length distribution and (d) generated token distribution by iteration type. A median agentic request issues about 2 LLM iterations, one intermediate and one final, while in the tail, the iteration depth goes up to 7. Similarly, the per iteration tool call fan out is 2 in the median case, but some iterations could have as high as 21 tool calls made in a single iteration. The median prompt length across intermediate and final iterations is about 20K tokens, while the tail could be 3x higher. The majority of the context is the system prompt (that could vary depending on the iteration and tools invoked in the previous iteration), and tool-specific instructions and formatting metadata. However, the median generated tokens for intermediate iterations is about 5x lower than the final iteration; this makes the intermediate iterations more prefill-bound, while the final iteration is decode-bound.

\subsection{Trace-driven analysis}
\label{sec:findings}

We present our key findings on the production workload.

\jheading{Finding 1 : Tool call execution dominates the tail FTR}
Figure~\ref{fig:trace_analysis}(e) presents the cumulative distribution of tool execution time as a fraction of total FTR latency across our workload. While the median request spends only 32\% of its FTR latency executing tools (with the remaining 68\% spent in LLM prefill and decode), the tail exhibits dramatically different behavior. At the 90th percentile, tool execution accounts for 61\% of FTR latency, climbing to 85\% at the 99th percentile. This tail dominance contradicts the conventional assumption that tool calls are lightweight I/O operations contributing marginally to end-to-end latency~\cite{autellix, continuum}.



Our analysis further reveals variability in tool execution latency. Figure~\ref{fig:trace_analysis}(f) shows box plots of tool latencies normalized to each tool’s median (p50) for six representative production tools. Even after normalization, all tools exhibit wide dispersion and pronounced right tails. 
By p75, normalized latency reaches 1.23–1.52× p50 across tools, and at p90 spans 1.60–3.28× p50. This heavy-tailed behavior indicates that tool execution latency remains highly sensitive to factors such as query complexity, backend contention, and the volume of data accessed (e.g., in search-style tools). As a result, tool execution time remains inherently difficult to predict, making tool-execution-time–aware KV caching strategies impractical in practice~\cite{continuum}.

The magnitude of tool execution time presents a clear optimization opportunity. Even partially overlapping tool execution with subsequent LLM computation could yield significant latency reductions. However, exploiting this opportunity requires tight coordination between the orchestrator and LLM engine, which is a capability fundamentally absent in current black-box architectures where components communicate solely through request-response interfaces.

\jheading{Finding 2 : Sequential orchestration leaves substantial parallelism unexploited}
Current orchestrators enforce strict sequential execution within each agentic request: iteration $i$ completes LLM decode, all tool calls execute to completion, all tool outputs return, and only then does iteration
$i+1$ begin prefill. This pipeline unnecessarily serializes the three phases -- prefill, decode, and tool execution that could potentially overlap. We identify two specific sources of unexploited parallelism that stem from this rigid sequencing.

\input{fig-tex/prefill_split_cdf}

\jheadingita{Opportunity 1: Prefill-tool overlap} Figure~\ref{fig:prompt_split_cdf} analyzes prompt composition across iterations, measuring the fraction of each prompt consisting of tool-independent content versus tool-dependent outputs. We observe that 50--80\% of iteration $i+1$'s prompt is available when iteration $i$ finishes decode. System instructions, conversation history, and templates don't depend on tool outputs. Only the final 20-50\% needs tool results. This creates a natural split point. The orchestrator could start prefill on the part of the prompt independent of tool outputs while tools execute.
and
extend the prefill incrementally.

\jheadingita{Opportunity 2 : Decode-tool overlap} LLMs generate tool calls as JSON: [{"tool": "search", "query": "..."}, {"tool": "plot", "query": "..."}]. Current orchestrators wait for the entire array before dispatching tools. But once the first JSON object completes (closing \}), that tool can execute immediately, allowing for streaming tool execution as decodes progress.

\jheading{Finding 3 : KV cache thrashing lowers prefix reuse} Agentic requests exhibit significant context reuse both across iterations within a request, as well as across requests that make similar tool calls. Across iterations, conversation history accumulates, whereas system prompts and instruction templates to consume tool output depend on tool call made in the prior iteration; such system prompts could be shared across requests that make similar tool calls. Standard prefix caching mechanisms should enable high cache hit rates by reusing these shared prefixes. However, we observe systematic cache thrashing that severely degrades performance under concurrent agentic workload execution.

\input{fig-tex/analysis-thrashing}

\jheadingita{Root cause: Workload-agnostic LRU eviction} Popular serving platforms \cite{vllm_sosp, zheng2023sglang} employ Least Recently Used (LRU) eviction policies that treat all cached KV blocks identically, lacking awareness of agentic request structure. To understand the thrashing behavior, consider three concurrent agentic requests $R_1$, $R_2$, and $R_3$ 
shown in Figure~\ref{fig:thrashing}. Each request performs two iterations, where the first iteration generates tool calls (denoted $T_{i1..1x}$, $T_{i1..2y}$, $T_{i1..3z}$ for requests $R_i$) and the second iteration can only begin after the previous iteration's tool execution completes.

At time $t=T$, all three requests complete their first LLM calls ($R_{11}$, $R_{21}$, $R_{31}$) and their KV blocks fill the available cache capacity in LRU order (Figure~\ref{fig:thrashing}(a)). When $R_2$'s second iteration ($R_{22}$) arrives at $t=T'$, it gets a prefix cache hit from $R_{21}$. However, inserting $R_{22}$'s new KV blocks causes LRU to evict $R_{11}$'s context (Figure~\ref{fig:thrashing}(b)). Subsequently, when $R_{12}$ arrives at $t=T''$, it must recompute the evicted blocks from $R_{11}$, which in turn triggers eviction of $R_{31}$'s context (Figure~\ref{fig:thrashing}(c)). This cascading pattern continues as $R_{32}$ arrives at $t=T'''$ and must recompute $R_{31}$'s evicted blocks, further evicting $R_{22}$'s context (Figure~\ref{fig:thrashing}(d)). 
The problem is that LRU evicts by recency, ignoring that $R_{11}$, $R_{21}$, $R_{31}$  are first-iteration contexts that will be reused by blocked second iterations. We address this by using workload-aware eviction hints to evict KV-cache regions with lower reuse likelihood \sref{sec:design:eviction_policy}.


In this work, we assume only HBM caching; thus, evicted KV blocks are recomputed. Alternatively, these blocks could be offloaded to secondary storage (e.g., CPU) using systems such as LMCache~\cite{lmcache} or Mooncake~\cite{mooncake}. In that case, our ideas remain applicable, as prefetching the right KV blocks is still required due to the HBM–storage latency hierarchy.

\subsubsection{Summary}
Our three key findings reveal a fundamental architectural problem: current agentic systems suffer from a \textbf{lack of co-design between the orchestrator and LLM inference engine}. Today's architectures treat these components as independent black boxes that communicate solely through opaque request-response interfaces. The orchestrator dispatches individual LLM calls and waits for complete responses before proceeding, while the engine treats each request as an isolated inference job with no awareness of its role in a multi-iteration agentic workflow. This decoupling prevents both layers from exploiting information that could enable critical optimizations.
\input{tables/tbl_codesign_api}

%% file: fig-tex/combined-analysis.tex
\begin{figure*}[t]
    \centering
    \includegraphics[width=\textwidth]{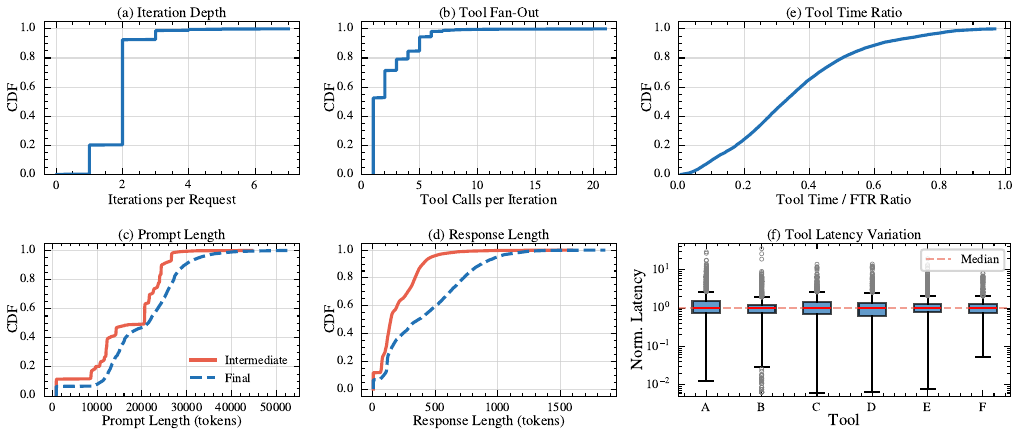}
    \caption{\small Detailed statistics of the agentic production trace. The distributions illustrate the structural characteristics of multi-step requests (a-d) and the associated tool execution dynamics (e-f). (e) denotes the normalized tool latency with respect to the median. 
    }
    \label{fig:trace_analysis}
\end{figure*}

%% file: fig-tex/prefill_split_cdf.tex
\begin{figure}[t]
    \centering
\includegraphics[width=0.75\linewidth]{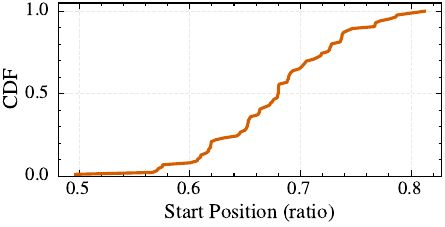}
\vspace{-0.7em}
    \caption{\small CDF of prompt prefix fraction independent of tool output}
    \label{fig:prompt_split_cdf}
\vspace{-0.7em}
\end{figure}

%% file: fig-tex/analysis-thrashing.tex
\begin{figure}[t]
    \centering
\includegraphics[width=0.95\linewidth]{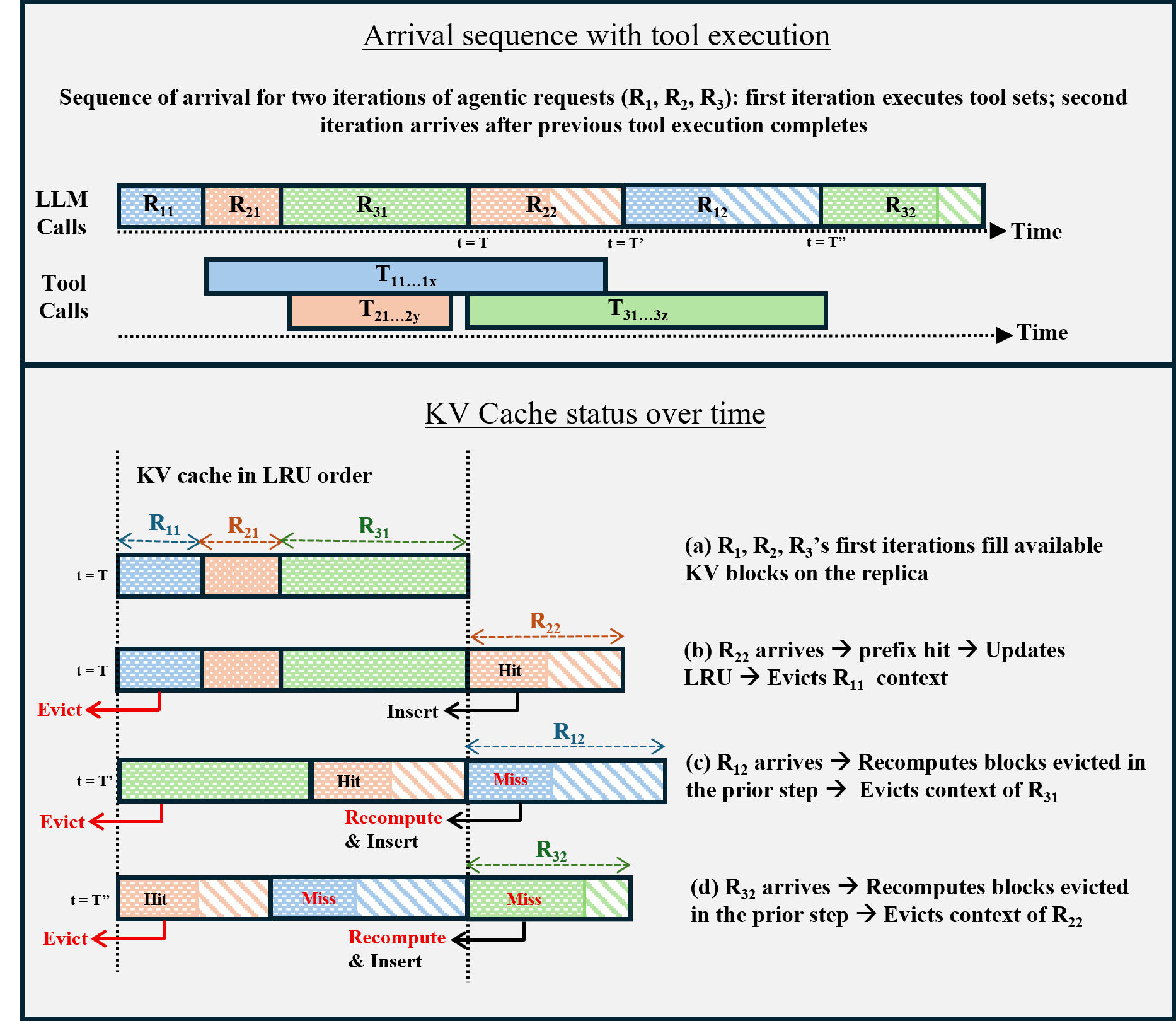}
\caption{\small Thrashing due to workload-agnostic KV eviction policy}
    \label{fig:thrashing}

\end{figure}

%% file: tables/tbl_codesign_api.tex

\begin{table}[t]
\centering
\footnotesize 
\scalebox{1.00}{
\begin{tabular}{p{3cm}p{5cm}}
\toprule
\textbf{API Call} & \textbf{Purpose} \\
\midrule
\texttt{submit\_partial\_} \texttt{prefill()} & Submit tool-independent prompt slice \\
\midrule
\texttt{extend\_prefill()} & Append tool outputs to pinned partial prefill context \\
\midrule
\texttt{register\_streaming\_} \texttt{callback()} & Receive partial decode outputs token-by-token \\
\midrule
\texttt{tag\_kv\_blocks()} & Annotate cached KV blocks with semantic hints (e.g., \texttt{system\_prompt}, \texttt{response}) \\
\midrule
\texttt{set\_reuse\_priority()} & Set priorities among KV blocks for pinning \\
\bottomrule
\end{tabular}}
\caption{\small APIs that enable orchestrator-engine co-design.}
\label{table:codesign_api}
\end{table}

%% file: sections/design.tex
\section{\sysname : Design and Implementation}
\label{sec:design}



\sysname extends the standard LLM serving architecture with a thin coordination layer that enables the orchestrator to communicate semantic hints about agentic request structure to the engine. 
As before, the orchestrator maintains knowledge of iteration boundaries, prompt composition, and tool dependencies, while the engine controls scheduling, batching, and KV cache management. However, through five new API calls shown in Table~\ref{table:codesign_api}, the orchestrator guides engine decisions without requiring model-level modifications.



\subsection{Parallel execution via prompt splitting}
Our analysis revealed that 50-80\% of iteration
$i+1$'s prompt is available immediately when iteration 
$i$ completes decode, yet current systems wait for all tool outputs before beginning any prefill computation. This sequential execution leaves  parallelism unexploited. However, naively starting prefill before tool outputs arrive introduces two challenges: (1) correctness: the engine must know where to splice tool outputs into the partial prompt, and (2) efficiency: the engine must retain the prefilled KV cache while tools execute, potentially without premature eviction.

Figure~\ref{fig:design-parallel-exec} illustrates our approach across three iterations of an agentic request. Each iteration consists of prefill ($P_i$) and decodes ($D_i$), followed by tool executions ($T_{ij}$ for tool $j$ in iteration $i$). Figure~\ref{fig:design-parallel-exec}a is the baseline sequential execution done by systems today; iteration 1 completes decodes $D_1$, all the tool executions $T_{11}, T_{12}$ and then iteration 2 begins prefill $P_2$. This pattern repeats for subsequent iterations. The LLM engine schedules iterations from other agentic requests in the idle time between tool calls to keep GPU occupied, creating long intra-request sequential chains of execution.

\input{fig-tex/design-parallel-exec}

 Prompt splitting as demonstrated in Figure~\ref{fig:design-parallel-exec}b breaks this sequential dependency by partitioning prompts into tool-independent and tool-dependent slices:
 \begin{enumerate}[noitemsep, leftmargin=*]
     \item \textbf{Slice identification: } When the iteration $i$
completes decodes and generates tool calls, the orchestrator 
aware of prompt template, identifies the insertion point where tool outputs will be spliced,  typically between system instructions/history and the tool results section.
\item \textbf{Eager prefill execution}: The orchestrator submits the tool-independent prefix $P\_{2a}$ using
\textit{submit\_\allowbreak partial\_\allowbreak prefill()}, which returns a continuation handle. The engine computes prefill while tools execute concurrently. The KV cache blocks from partial prefill are tagged with high  priority via $set\_reuse\_priority()$ to prevent eviction.
\item \textbf{Prompt extension:} Once tool outputs arrive, the orchestrator constructs the tool-dependent suffix and calls $extend\_prefill()$ with the continuation handle. The engine splices the new content ($P_{2b}$) onto the pinned KV cache from $P_{2a}$, completes the prefill, and proceeds to decode $D_2$.  If tool execution fails or times out, the orchestrator 
discards the partial prefill and sets appropriate hints for its KV for the engine to release pinned resources.
 \end{enumerate}
 This approach overlaps tool execution with a part of prefill, reducing the request's end-to-end latency.

\subsection{Streaming tool dispatch with decodes}
Agentic systems generate tool calls as structured JSON arrays during the LLM decode phase. Figure~\ref{fig:design-parallel-exec}c shows that current systems wait for complete decode output before parsing and dispatching tools, introducing unnecessary serialization. However, once a complete tool invocation structure is decoded (e.g., the closing \} of ${"tool": "search", "query": "..."}$), that tool can execute immediately.

\sysname implements streaming tool dispatch through token-level callbacks from the engine to the orchestrator. When submitting a decode request that will generate tool calls (intermediate iterations), the orchestrator calls $register\_\allowbreak streaming\_\allowbreak callback()$ with a handler function. The engine invokes this handler after each decoded token. The orchestrator maintains a streaming JSON parser that accumulates tokens and identifies complete objects. When a tool call object closes (final \} token), the parser extracts the tool name and parameters. As soon as a complete tool invocation is identified, the orchestrator dispatches it for execution without waiting for remaining decode tokens. 
Subsequent tool calls in the JSON array are dispatched as they complete. 
The orchestrator tracks completion of all tools before proceeding to the next iteration.

\subsection{Workload-aware KV cache management and scheduling}
\label{sec:design:eviction_policy}
Our analysis  identified systematic KV cache thrashing in concurrent agentic workloads despite significant context reuse. Figure~\ref{fig:thrashing} illustrates the root cause: workload-agnostic LRU eviction treats all cached blocks identically, causing cascading evictions that destroy reuse opportunities. LRU evicts based solely on recency, ignoring workload structure. It cannot distinguish between low-value blocks (transient content unlikely to be reused) and high-value blocks (first-iteration contexts that will be reused by second iterations currently blocked on tool execution). In \sysname, we use semantic tagging and priority-based eviction to tackle this. 

\jheading{KV block tagging} When the orchestrator submits LLM requests, it tags KV cache blocks with semantic metadata using $tag\_kv\_blocks()$.
\begin{tightitemize}
    \item \textbf{System prompts:} Marked as $SYSTEM\_PROMPT$ with high reuse priority. Depending on the iteration type (intermediate or final) and the combination of tool calls made in the previous iteration, there could be many variants of system prompts that the orchestrator constructs using predefined rules; for e.g., the system prompt could contain instructions on how the output of the tool call made in the previous iteration should be consumed and interpreted. Such system prompts occur before the user-specific query and hence could be  shared across many requests.
    \item \textbf{User specific query:} Marked as $USER\_QUERY$ and contains the request-specific context including user information which can result in prefix matches intra-request alone; across iterations of the same agentic request. 
    \item \textbf{Tool outputs}: tagged as $TOOL\_OUTPUT\_ITER\_i$, these are reused by subsequent intermediate iterations in an agentic request, provided that these iterations share the system prompt. If the system prompt in the next iteration changes due to a new tool call in the prior iteration, the prefilled tool output context from a prior iteration becomes useless as it is appended after the system prompt and user query.
    \item \textbf{Final response}: Tagged as $RESPONSE$ The final LLM call's decodes are the user-facing output tokens; these have no reuse potential and are candidates for eviction.
    \item \textbf{Partial prefills}: These are KV blocks from \textit{submit\_\allowbreak partial\_\allowbreak prefill()}, tagged as $PARTIAL\_PREFILL$ with maximum priority until its extension prompt is completed. 
\end{tightitemize}

\jheading{Priority-based eviction policy}. Using the tags on the KV blocks, the engine's scheduler implements a priority-aware eviction policy that respects orchestrator hints. When cache capacity is exceeded, the scheduler evicts the lowest-priority blocks first. Within a priority tier, LRU ordering is used as a tiebreaker. 
The eviction priority (evicted first) -> (evicted last) is as follows: 


\[
\scalebox{0.8}{$
\begin{aligned}
\text{RESPONSE} \;\longrightarrow\;
\text{TOOL\_OUTPUT} \;\longrightarrow\;
\text{USER\_QUERY} \\
\;\longrightarrow\;  \text{SYSTEM\_PROMPT} \;\longrightarrow\;
\text{PARTIAL\_PREFILL}
\end{aligned}
$}
\]

\input{fig-tex/no-thrashing}
 \jheading{Example} Returning to the scenario in Figure~\ref{fig:thrashing},  our policy prevents cascading evictions as shown in Figure~\ref{fig:no-thrashing}. At time $t=T'$, when $R_{22}$ needs to insert new blocks, the eviction policy targets low-priority blocks (tool outputs) rather than $R_{11}$'s context, which is tagged with high reuse priority since $R_{11}'s$ tools are executing. When $R_{12}$ arrives at $t=T"$, it finds $R_{11}'s$ blocks still cached, achieving a prefix hit and avoiding recomputation. Similarly,  $R_{31}$ survives until $R_{32}$ arrives, preventing the cache miss chain.

 \jheading{Workload-aware scheduling}. 
 While scheduling is orthogonal to \sysname{}'s optimizations, it remains a critical component for efficient agentic systems. Although agentic requests typically issue multiple LLM calls, the existing serving engines schedule at the granularity of individual LLM calls (e.g., FIFO), without global agentic context.
As a result, requests that issue LLM calls more frequently may be unfairly prioritized over earlier-arriving agentic requests. To address this, \sysname{} enforces a workload-aware policy in the serving engines that preserves a global FIFO ordering of LLM calls with respect to the arrival time of agentic requests in the orchestration, ensuring fairness. 
Integrating more sophisticated scheduling techniques designed to prevent starvation, such as Autellix\cite{autellix}, is left for future work.

\subsection{Implementation}
We implement \sysname{} using $\approx$3,500 lines of Python consisting of 
a serving engine layer and an independent orchestrator. 
We build the serving engine on top of vLLM (v0.11.0) and enable chunked prefill  with prefix caching by default.
For the orchestrator, we developed a lightweight, asyncio-based event-driven framework designed to accurately replay synthetic production traces. \sysname{}'s modular design readily supports integration with alternative serving backends \cite{tensorrt, deepspeed}. 
Our engine modifications natively embed workload semantics, allowing the orchestrator to leverage this contextual metadata. For robustness, the components rely on a heartbeat-based membership protocol. 


%% file: fig-tex/design-parallel-exec.tex
\begin{figure}[t]
    \centering
\includegraphics[width=0.95\linewidth]{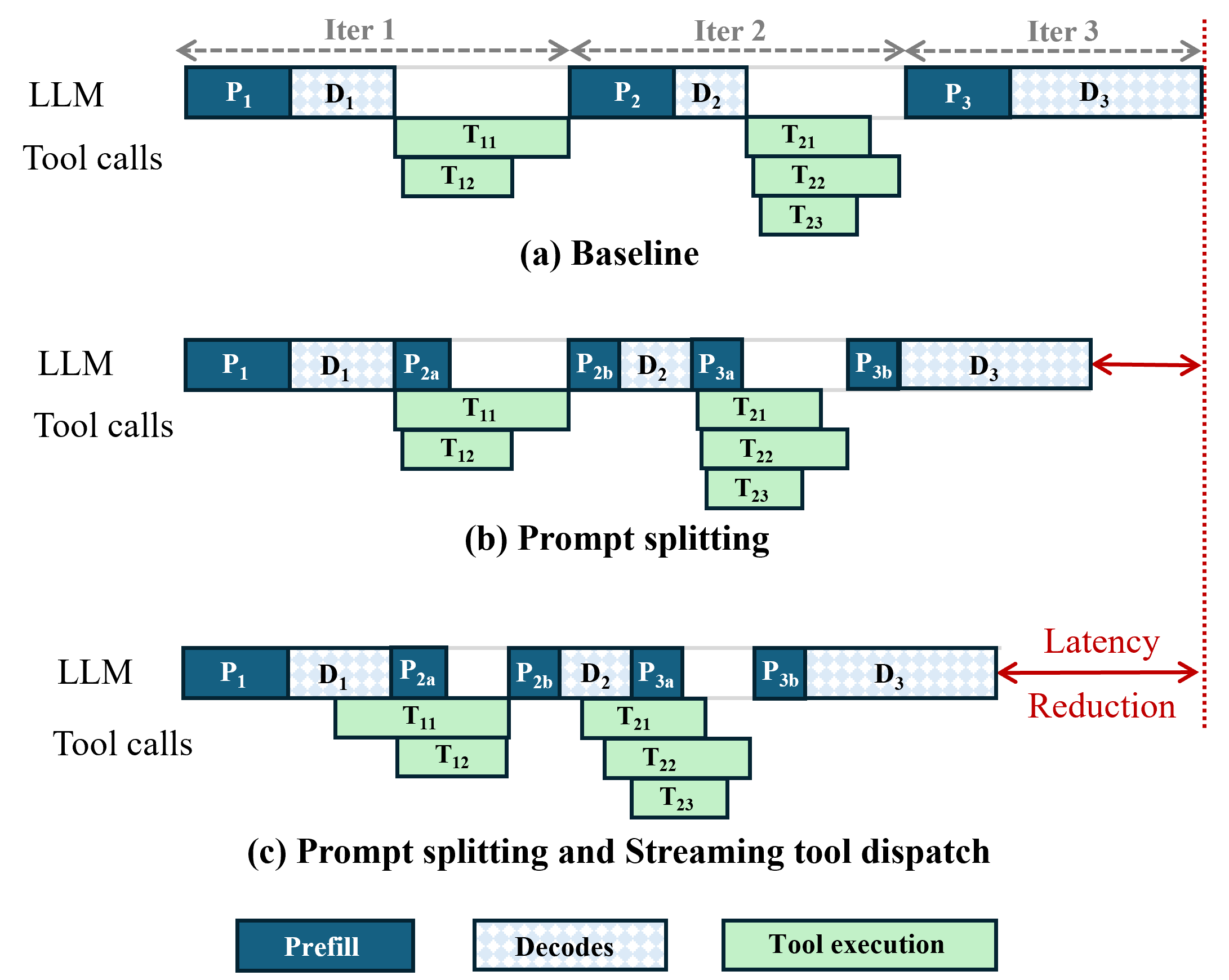}
\caption{\small Intra-request parallel execution in \sysname}
\label{fig:design-parallel-exec}
\end{figure}

%% file: fig-tex/no-thrashing.tex
\begin{figure}[t]
    \centering
\includegraphics[width=\linewidth]{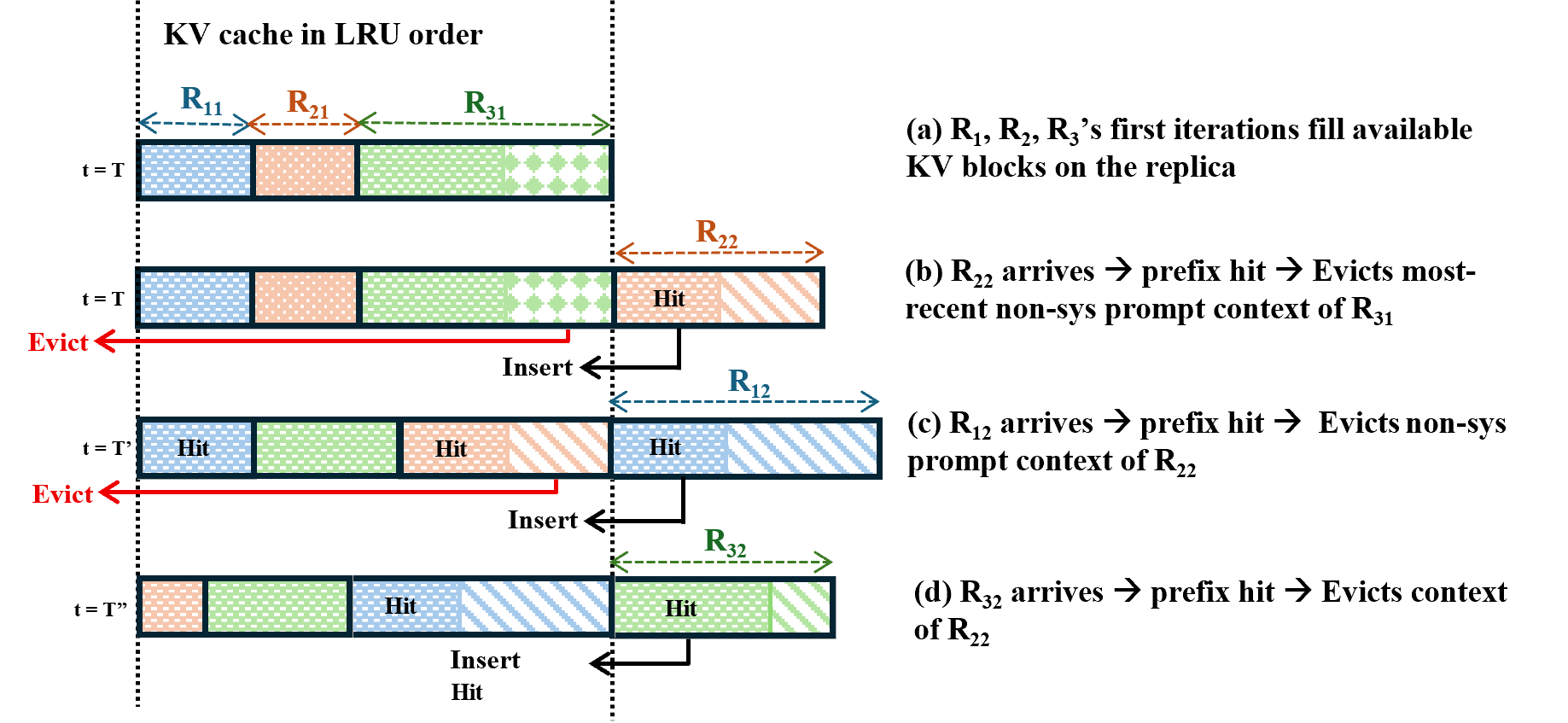}
\vspace{-2em}
\caption{\small Workload-aware KV eviction policy}
    \label{fig:no-thrashing}
\end{figure}

%% file: sections/eval.tex
\input{fig-tex/eval-load-vs-latency}
\section{Evaluation}
\label{sec:eval}

Our evaluation answers the following questions:
\begin{enumerate}
    \item Does \sysname achieve a better throughput-latency trade-off  for FTR and E2E latency on production agentic workloads?
    \item Can \sysname generalize to disaggregated serving and diverse LLMs?
    \item How does \sysname perform on open source agentic datasets?
    \item What is the independent contribution of each optimization in \sysname?
    \item Does \sysname increase KV cache hit rates?
\end{enumerate}

\subsection{Experimental Setup}
\jheading{Models}
We evaluate \sysname on Qwen3-14B~\cite{qwen3}, a 14B-parameter open-source model featuring native function-calling support. At BF16 precision, this model fits within the memory budget of a single A100-80GB GPU supporting a 128K context window and accommodates the maximum prompt lengths observed in our workloads. 
\sysname's optimizations operate at the orchestrator-engine interface independent of the models or kernels used in the engine.
To validate this, we evaluate \sysname on Gemma-27B in \S\ref{sec:eval:generalizability}.

\jheading{Workloads}
We evaluate \sysname{} across production and open-source traces.

\jheadingita{Production Trace} 
It uses a subset of the production trace detailed in \S\ref{sec:analysis}. 
We pick a stratified sample of 120 requests from the production trace, characterized by a high degree of tool-call fan-outs per request. We replay it using the original prompts, recorded tool outputs, and the decode lengths extracted from the trace.  Tool-call latencies are normalized according to the observed tool-to-
LLM ratio within the original traces as described in \S\ref{sec:analysis}. Additionally, we present the evaluation of \sysname on a different subset of production traces with median tool-call fan-out in Appendix and observe similar magnitude of gains.
Request arrival times follow a Poisson process.
We report the mean success metrics with confidence bands across five  seeds.

\jheadingita{Open-Source Traces}
The traces are derived from the following open-source benchmarks---

\begin{tightitemize}
  \item \textbf{BFCL v4 Web Search~\cite{bfcl}:}
  An open-source agentic benchmark that evaluates multi-hop web search via function-calling agents with standardized tools.
  We use 56 single-turn requests randomly sampled from an initial set of 100, each containing at least one multi-tool invocation.
  Agent trajectories are short, with a mean of 4.23 iterations and low tool fan-out per iteration ($\approx 2$).
  Tool execution dominates latency, averaging 1.09\,s per call (variance 1.7\,s) and accounting for approximately 40\% of E2E request time.

  \item \textbf{SWE-Bench~\cite{swebench}:}
  An open-source agentic software engineering benchmark that captures long-horizon code exploration, execution, and repair over real-world GitHub issues.
  We evaluate 60 single-turn requests randomly sampled from a pool of 500, exhibiting substantially longer agent trajectories (mean 20.0 iterations) with similarly low tool fan-out ($\approx 2$).
  Tool calls are shorter on average (0.29\,s, variance 1.14\,s) and contribute roughly 30\% of overall execution time.
\end{tightitemize}

Both traces follow a strictly append-only prompt structure, where each iteration appends new context without modifying prior prompt prefixes.
Most evaluations use the production trace, unless explicitly stated. 

\jheading{Hardware}
We evaluate \sysname on A100-80GB GPUs \cite{a100azure}. For collocated experiments, we use a single A100. For disaggregated experiments, we use two A100-80GB GPUs - one dedicated prefill node and one dedicated decode node - connected over NVLink. 

\jheading{Baselines}
We compare against a baseline using vLLM as the serving backend with Prefill-Decode co-location at a chunk size of 256, utilizing request-aware scheduling that maintains FIFO ordering at an agentic request level instead of individual LLM calls.
\sysname builds on this baseline and additionally introduces prompt splitting, streaming dispatch, and orchestrator-aware KV cache management. 
 We also compare against the PD disaggregated baseline in vLLM. 
Automatic prefix caching is enabled by default. 
We also compare against Continuum~\cite{continuum}, a concurrent work that optimizes agentic systems by pinning KV cache blocks during tool execution using a time-to-live mechanism to prevent eviction-induced recomputation across iterations. We integrate Continuum with our orchestrator on a best-effort basis and 
set the TTL to 6s, based on mean tool call time in our production trace.

\jheading{Metrics}
We evaluate \sysname{} on two primary latency metrics. \textit{First Token Rendered} (FTR) measures the time from the initial user request submission to the generation of the first token in the final user-visible response. 
\textit{End-to-End} (E2E) latency reflects the total request completion time.
\input{fig-tex/generalizibility}

\subsection{End-to-End Performance}
\label{sec:eval:e2e}
We assess whether \sysname achieves a better throughput-latency tradeoff compared to baselines. Specifically, we evaluate \sysname{} on varying loads (0.0075, 0.01, 0.0125 and 0.0015 QPS) and measure p50/p90 FTR and E2E latency. 
To characterize this trade-off, we plot serving capacity curves with ingest load (QPS) on the y-axis and latency on the x-axis. A system exhibits a better throughput–latency trade-off if it can sustain higher load at lower latency, or equivalently, achieve lower latency at the same load—corresponding to curves that lie closer to the top-left of the plot.

Figure~\ref{fig:e2e} shows the serving capacity curves for both \sysname{} and baseline. \sysname{} achieves a better throu\allowbreak -ghput-latency tradeoff compared to baseline,  with its service capacity curve lying to the left of the baseline. At the same p50 FTR latency, \sysname{} sustains up to 77\% higher load; conversely, for a fixed load, it achieves up to a 15\% reduction in p50 FTR latency. \sysname{} also delivers improved trade-offs for tail latency. For the same p90 FTR latency, it sustains up to 45\% higher load at the same latency, or achieves up to an 11\% reduction in tail FTR latency at the same load. Similar benefits extend to end-to-end (E2E) latency, with up to a 9\% reduction in p90 E2E latency. \sysname{} achieves a better trade-off as it unlocks intra-request parallelism through its co-design and hides tool call time under prefills and decodes.

The gains are more pronounced for FTR than for E2E latency because the E2E metric includes the decode phase of the final iteration, which does not trigger any subsequent tool execution. As a result, this phase cannot benefit from prompt splitting or streaming dispatch optimizations, limiting the achievable acceleration in E2E latency. To understand these gains, we show detailed FTR and E2E latency CDFs across varying ingest loads in Appendix (Figure \ref{fig:eval:production_trace}, \ref{appendix:gpt4-bars}).
\subsection{Generalizability}
\label{sec:eval:generalizability}

\jheading{Disaggregated Serving} 
We evaluate \sysname{} in disaggregated prefill-decode setting to assess if \sysname{}'s generalizes to a different deployment.
We run these workloads at 0.015 and 0.025 QPS, maintaining a per-GPU load equivalent to our collocated experiments.
Figure~\ref{fig:generalizability} (a) summarizes the FTR and E2E latency improvements achieved under disaggregated serving. Across both evaluated load levels, \sysname{} consistently reduces both median and P90 FTR by 12\% to 16\%. 
These gains are consistent with the collocated results, demonstrating that \sysname{}'s optimizations transfer across deployment topologies without modification.

\jheading{Model Generalizability} \sysname{}'s optimizations operate entirely at the orchestrator-engine interface 
and universally apply to tool-calling LLMs served through vLLM. To validate this architecture-agnostic claim, we evaluate our system using Gemma-3-27B~\cite{gemma3}: a more parameter-heavy model with a different attention mechanism than Qwen3-14B. 



Figure~\ref{fig:generalizability} (b) shows the median FTR and E2E improvements for both models at 0.0075 QPS. \sysname{} reduces median FTR by 13.3\% and E2E latency by 9.2\% on Gemma-3-27B. 
This demonstrates \sysname{}'s latency improvements generalize across different models. 
The performance difference between the two models directly stems from Gemma-3-27B's higher LLM compute requirements. Specifically, its prefill and decode phases are 1.34$\times$ and 1.60$\times$ slower, respectively. This increased compute time  reduces the tool fraction of the overall FTR from 12.9\% down to 9.0\% 
limiting the available time window for \sysname{}'s optimizations to take effect.

\jheading{Open Source Traces} 
We evaluate \sysname{} on the BFCL Web Search and SWE-Bench traces 
using the Qwen3 \allowdisplaybreaks ‑14B model. Table~\ref{fig:generalizability} (c) shows the results across varying load levels. The baseline saturates at different load levels for both traces (2 QPS for BFCL and 0.25 QPS for SWE-Bench).
\sysname{}'s optimizations generalize and reduce median FTR for both the traces.
\sysname{} causes 10\% and 13.2\% p50 FTR reduction at 1.5 QPS load on BFCL, and 0.05 QPS load on SWE-Bench respectively.
The lower median improvement observed on the open-source traces relative to the production trace (7.2-13.2\% versus 15.3\%) 
is primarily due to two key differences, the BFCL and SWE-Bench traces have a) less tool call fan-out that limits the opportunities to overlap tool calls with the decode, and b) strictly append-only prompt structure that limits gains from prefill splits as the splitted prefill may already remain prefix cached (from previous iteration's prefill) leading to less overlap with tool calls.

\subsection{Ablation Study}
\label{sec:eval:ablation}
\subsubsection{Cumulative Impact of Optimizations} Table~\ref{tbl:ablation} isolates the contribution of each optimization at 0.0075 QPS by incrementally adding them.


\jheading{Prompt Splitting}
Prompt splitting reduces median FTR by 6.1\% and E2E latency by 3.5\%. 
Although 50--80\% of the prompt is available for the split (\S\ref{sec:findings}), these time savings are bounded the extent of overlap between prefill and  tools calls.
For instance, whenever a tool executes faster than the partial prefill, the remaining prefill execution still remains on the critical path. Moreover, E2E latency gains remain less compared to FTR as decode dominates E2E latency and this optimization only performs prefill overlaps with tool calls. 
\input{tables/eval-ablation}

\jheading{Streaming Dispatch}
Streaming dispatch further contributes to latency reduction by overlapping decodes with tool calls. It achieves an 8.3\% FTR and 5.5\% E2E latency improvement on top of prompt splitting, yielding cumulative improvements of 14.4\% and 9.0\%, respectively. 
Especially for high tool call fan-out requests, this optimization converts a purely sequential decode-then-dispatch pipeline into overlapped execution reducing the critical path.

\jheading{KV Cache Management}
This optimization further contributes a 1.8\% improvement in both FTR and E2E latency. This reflects a direct reduction in prefill recomputation driven by higher cache hit rates 
under concurrent agentic workloads.
as demonstrated in \S\ref{sec:eval:kv}.
However, these increased hit rates  don't proportionally translate to similar latency gains  
as the decode dominates the E2E latency.

\input{fig-tex/top5-ftr}
\subsubsection{Per-Request FTR Latency Breakdown}
To isolate the sources of performance improvement, Figure~\ref{fig:latency_breakdown} decomposes the FTR latency of five randomly selected tool-heavy requests  
into three distinct phases: critical path tool time, total prefill time, and total decode time summed across intermediate iterations of each request. The breakdown reveals two key advantages. First, \sysname{} reduces the critical path tool time 
which is a direct consequence of streaming tool executions concurrently with the decode. 
Second, \sysname{} decreases the total prefill time compared to the baseline. This is due to higher cache hit rates \sref{sec:eval:kv} and 
overlapping prefill execution with the tool invocations. 
The breakdown reveals how \sysname{} reduces FTR latency and enables better throughput-latency trade-offs \sref{sec:eval:e2e}. 

\subsubsection{KV Cache Analysis}
\label{sec:eval:kv}

\input{fig-tex/eval-kv-analysis}

Figure~\ref{fig:eval:kv} illustrates the decomposition of inter-request versus intra-request cache hits across varying iteration depths of requests. At depth~0, cache hits are predominantly inter-request as requests share system prompt prefixes.
\sysname{}'s priority policy explicitly prevents these prefixes from eviction leading to higher cache hits. 
As iteration depth increases, context grows with accumulation of previous iterations' tool call outputs.
\sysname{} increases intra-request hits as partial prefills issued by \sysname{} contains tool call outputs from previous iterations and are given least eviction priority.
Overall, \sysname{} improves the global cache hit rate from 21.8\% to 44.6\%. 

\section{Comparison with Concurrent Works}
\label{eval:continuum}

Figure~\ref{fig:continuum} compares \sysname against Continuum~\cite{continuum} configured with TTL~=~6s (average tool execution time in production trace). \sysname reduces median FTR by 17\% over Continuum. Continuum addresses only the cache thrashing bottleneck and does not overlap tool execution with prefill computation nor dispatches tools during decode. The execution in Continuum remains sequential 
regardless of TTL configuration. Furthermore, TTL-based pinning is sensitive to tool execution variance. A higher variance leads to  misestimations of optimal TTL which mitigates benefits of KV pinning. Our trace shows a high variation in tool call times (\S\ref{sec:analysis}), meaning a fixed TTL of 6s may cause cache evictions leading to higher tail latency. 
\sysname's semantically aware priority based cache eviction policy remains effective regardless of tool execution variance.

%% file: fig-tex/eval-load-vs-latency.tex
\begin{figure*}[!t]
    \centering
    \includegraphics[width=\textwidth]{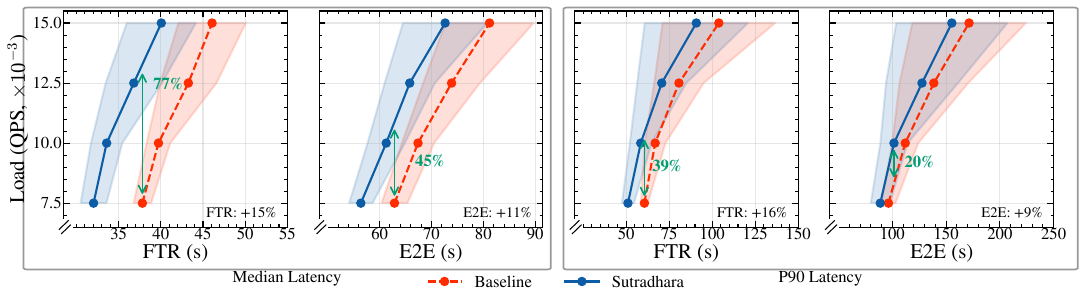}
    \caption{ \small Serving capacity curves for \sysname and Baseline with p50/p90 FTR and E2E metrics on x-axis and ingest load (QPS) on y-axis. Curves that lie closer to the top-left are better: higher load sustained at lower latency. \sysname{} is closer to the top-left with a better throughput-latency tradeoff compared to baseline: sustains upto 77\% higher load at the same latency or achieves upto 15\% reduction in p50 FTR latency at the same load. Similar gains extend to E2E latency.} 
    \label{fig:e2e}
\end{figure*}

%% file: fig-tex/generalizibility.tex
\begin{figure*}[t!]
    \centering
    \begin{minipage}[b]{0.30\linewidth}
        \centering
        \includegraphics[width=0.9\linewidth]{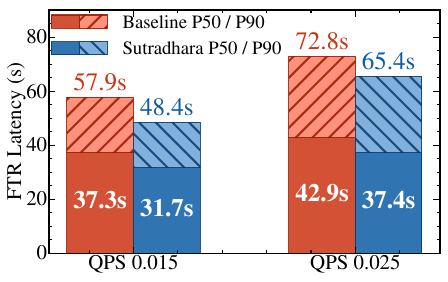}
        \vspace{-1em}
        \caption*{\small (a) Disaggregated serving}
    \end{minipage}
    \begin{minipage}[b]{0.30\linewidth}
        \centering
        \includegraphics[width=0.84\linewidth]{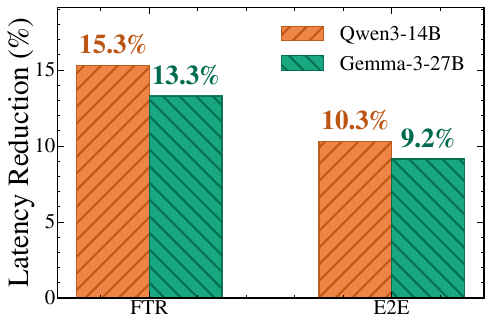}
        \vspace{-1em}
        \caption*{\small (b) Model generalizability}
    \end{minipage}
    \begin{minipage}[b]{0.35\linewidth}
        \centering
        \small
        \begin{tabular}{clrrc}
    \toprule
    & QPS & BL (s) & \sysnameshort{} (s) & Gain \\
    \midrule
    \multirow{3}{*}{\rotatebox[origin=c]{90}{BFCL v4}} 
    & 0.50 & 20.37 & 18.91 & 7.2\% \\
    & 1.50 & 20.87 & 18.79 & 10.0\% \\
    & 2.00 & 21.59 & 19.87 & 8.0\% \\
    \midrule
    \multirow{3}{*}{\rotatebox[origin=c]{90}{\shortstack{SWE \\ Agent}}} 
    & 0.05 & 48.47 & 42.10 & 13.2\% \\
    & 0.10 & 51.95 & 45.20 & 13.0\% \\
    & 0.25 & 61.86 & 56.77 & 8.2\% \\
    \bottomrule
\end{tabular}
        \vspace{-3mm}
        \caption*{\small (c) Median FTR on open source traces}
    \end{minipage}
    \vspace{-1em}
    \caption{\small \sysname{} latency gains generalize to (a) disaggregated serving, (b) different models (Gemma3-27B), and (c) open source traces (BL = Baseline and SD = Sutradhara)}

    \label{fig:generalizability}
\end{figure*}

%% file: tables/eval-ablation.tex
\begin{table}[t!]
\centering
\caption{\small Each optimization's contribution at 0.0075 QPS. The median latency averaged over 3 seeds. PS: prompt splitting, DS: streaming dispatch, KV: cache management. Improvements are relative to Baseline+Sched.}
\label{tbl:ablation}
\resizebox{\columnwidth}{!}{%
\begin{tabular}{lrrrrrrr}
\toprule
Config & FTR (s) & E2E (s) & \multicolumn{2}{c}{Cumulative} & \multicolumn{2}{c}{Incremental} \\
\cmidrule(lr){4-5} \cmidrule(lr){6-7}
 & & & FTR & E2E & FTR & E2E \\
\midrule
Baseline & 37.45 & 61.57 & —       & —      & —      & —      \\
+PS            & 35.17 & 59.39 & +6.1\%  & +3.5\% & +6.1\% & +3.5\% \\
+PS+DS         & 32.07 & 56.03 & +14.4\% & +9.0\% & +8.3\% & +5.5\% \\
+PS+DS+KV      & 31.39 & 54.93 & +16.2\% & +10.8\% & +1.8\% & +1.8\% \\
\bottomrule
\end{tabular}}
\end{table}

%% file: fig-tex/top5-ftr.tex
\begin{figure}[t]
    \centering
    \includegraphics[width=0.9\columnwidth]{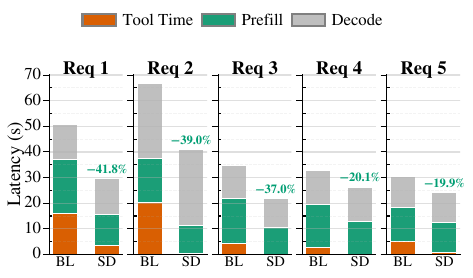}
    \caption{\small FTR Latency breakdown across five representative requests. B is baseline, \sysnameshort{} is \sysname{}}
    \label{fig:latency_breakdown}
\end{figure}

%% file: fig-tex/eval-kv-analysis.tex
\begin{figure}[t]
\centering
\includegraphics[width=0.9\columnwidth]{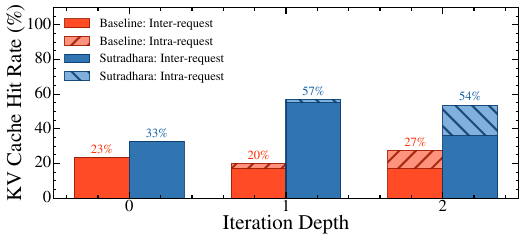}
\caption{\small \sysname{} vs baseline on inter- vs 
intra-request cache hit rate by agentic requests' iteration depth.}
\label{fig:eval:kv}
\end{figure}

%% file: sections/related.tex
\section{Related Work}
Table~\ref{tab:feature-comparison} compares \sysname{} with related works.

\jheading{Agent Development}
Several frameworks, such as Langraph \cite{langgraph}, Autogen \cite{autogen}, Dspy \cite{dspy}, and Palimpzest \cite{palimpzest}, support agentic development by providing orchestration strategies for different applications. However, these frameworks treat the LLM backend as a black box and focus solely on orchestrator design. In contrast, our proposed interface enables co-design between the orchestrator and the LLM backend, which can be leveraged by these frameworks to  improve system efficiency.

\input{fig-tex/continuum}

\jheading{Agentic System Optimization}
Prior works have explored optimizing agentic inference, but they differ fundamentally from \sysname{} in scope and assumptions. Parrot \cite{parrot} and Murakkab \cite{murakkab} introduce declarative APIs to represent agentic workflows as static DAGs, allowing the LLM backend to optimize throughput–latency trade-offs using DAG information. In contrast, the tool-based agentic applications targeted by \sysname{} are not statically defined, and their DAG structure is unknown beforehand, making such approaches inapplicable. Autellix \cite{autellix} proposes a non-clairvoyant scheduler to improve throughput, while KVFlow \cite{kvflow} focuses on KV-cache management for multi-agent applications. Circinus \cite{circinus} introduces SLO-aware query planning to improve goodput, and DroidSpeak \cite{droidspeak} explores KV-cache sharing across agents in multi-agent orchestration. However, these systems assume that agentic latency is not significantly influenced by tool calls—an assumption that does not hold in production, where tool calls are prevalent and contribute substantially to latency. \sysname{} addresses this gap by making agentic inference tool-aware through orchestrator–LLM backend co-design.
Continuum \cite{continuum} seeks to optimize KV cache management by predicting tool invocation times; however, our analysis reveals that these times exhibit substantial variability across requests, rendering accurate prediction impractical in our context. The closest work, Conveyor \cite{conveyor}, overlaps partial tool execution with decoding. However, partial execution is infeasible for the agentic applications considered in the paper, as many tools cannot execute with incomplete parameters. Instead, \sysname{} tackles the practical challenge of handling tool-call fan-out, overlapping them with decoding.

\input{tables/relatedworks}


%% file: fig-tex/continuum.tex
\begin{figure}[t]
    \centering
    \includegraphics[width=0.7\linewidth]{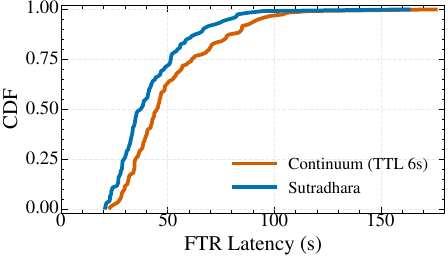}
    \caption{\small FTR-latency CDF for \sysname and Continuum.}
    \label{fig:continuum}
\end{figure}

%% file: tables/relatedworks.tex
\begin{table}[tb]
\centering
\caption{Comparison of related works against \sysname.  \cmark-supported; \lmark-limited support; \xmark-absent}
\label{tab:feature-comparison}
\setlength{\tabcolsep}{4pt}
\renewcommand{\arraystretch}{1.2}
\resizebox{\columnwidth}{!}{%
\begin{tabular}{@{}l*{5}{c}@{}}
\toprule
\textbf{Works}
    & \shortstack{\textbf{Dynamic}\\\textbf{Tool-Based}\\\textbf{Agents}}
  & \shortstack{\textbf{Workload-Aware}\\\textbf{KV Cache}\\\textbf{Management}}
  & \shortstack{\textbf{Tool--Decode}\\\textbf{Execution}\\\textbf{Overlap}}
  & \shortstack{\textbf{Prefil--}\\\textbf{Tool}\\\textbf{Overlap}} \\
\midrule
Parrot~\cite{parrot}
  & \xmark & \xmark & \xmark & \xmark \\
Murakkab~\cite{murakkab}
  & \xmark & \xmark & \xmark & \xmark \\
Autellix~\cite{autellix}
  & \cmark & \xmark & \xmark & \xmark \\
Continuum~\cite{continuum}
  & \cmark & \cmark & \xmark & \xmark \\
Conveyor~\cite{conveyor}
  & \cmark & \xmark & \lmark & \xmark \\
\midrule
\textbf{\sysname (Ours)}
  & \cmark & \cmark & \cmark & \cmark \\
\bottomrule
\end{tabular}%
}
\end{table}

%% file: sections/conclusion.tex
\section{Conclusion}

We present an extensive analysis of production tool-based LLM agentic applications, identifying three fundamental bottlenecks: (i) tool latency dominates end-to-end response time (30–85\%), (ii) KV-cache effectiveness degrades despite significant context reuse across iterations, and (iii) sequential orchestration under-utilizes intra-request parallelism. These inefficiencies stem from decoupling orchestration from the LLM backend. We address this with \sysname{}, a co-designed agentic serving system. \sysname{} enables prompt-level parallelism, overlaps tool-call fan-outs with decoding, and incorporates workload-aware KV-cache management and scheduling. \sysname{} sustains up to 77\% higher load at the same median latency, highlighting the need for coordinated orchestration–LLM co-design to scale agentic systems efficiently.

%% file: sections/appendix.tex
\setcounter{figure}{0}
\setcounter{table}{0}

\renewcommand{\thefigure}{\Alph{section}.\arabic{figure}}
\renewcommand{\thetable}{\Alph{section}.\arabic{table}}

\section{Appendix}

\input{fig-tex/eval_production_e2e_cdf}
\input{fig-tex/gpt4_60_bars}

\subsection*{A.1 Detailed Latency CDFs on Production Trace}
Figure~\ref{fig:eval:production_trace} shows the full FTR and E2E latency CDFs for \sysname and the baseline across three load levels, evaluated on a subset of the top 120 high fan-out requests from the production trace. At 0.0075 and 0.01 QPS, \sysname consistently shifts the CDF left across the entire distribution, yielding median improvements of 15--16\% in FTR and 9--10\% in E2E. Furthermore, at 0.0125 QPS---which is 25\% beyond the baseline's maximum sustainable load \sysname achieves a lower median FTR (37s versus 40s) and E2E (66s versus 67s) than the baseline operating at the strictly lower 0.01 QPS. This directly confirms the 25\% higher serving capacity reported in \S\ref{sec:eval:e2e}.

\subsection*{A.2 Robustness Across Production Trace Subsets}
\label{appendix:prod}
Figure~\ref{appendix:gpt4-bars} validates that \sysname's gains are not artifacts of the high fan-out subset used in the primary evaluation. We evaluate a separate 60-request subset sampled to represent the median tool call fan-out of the same production trace. Across all load levels, \sysname consistently reduces FTR P50 by 17--18\% and E2E P50 by 6--11\%, closely matching the performance gains reported on the primary trace. This confirms that \sysname's architectural optimizations remain robust across diverse operating regimes within the production workload distribution.

\subsection*{A.3 Limitations of  Orchestrator-Engine Coupling}
Frameworks like LangGraph~\cite{langgraph} are well positioned to adopt \sysname's interface. LangGraph's graph-based execution model exposes explicit boundaries between LLMs and tool execution as discrete named nodes. The key adaptation is extending the LLM node to support concurrent operations during decode i.e eagerly submitting tool-independent prompt content and dispatching individual tools as their specifications complete, rather than treating decode as a strictly sequential, atomic step. These changes are confined to the LLM node and the interface layer between the orchestrator and the engine, leaving LangGraph's core execution model intact. We expect the same interface to generalize to other graph-based frameworks such as AutoGen~\cite{autogen} as they adopt similar execution models.

While \sysname's co-design of the orchestrator and serving engine is strictly necessary to enable performance optimizations like streaming dispatch and orchestrator-aware KV cache pinning, it inherently trades deployment flexibility for execution efficiency. Specifically, \sysname's extended five-API interface introduces a tight versioning dependency between the orchestrator and the LLM serving engine. Upgrading either component requires verifying API compatibility, adding operational overhead compared to traditional black-box deployments where the two layers evolve independently. Consequently, organizations adopting \sysname must coordinate releases across both layers rather than treating the inference engine as a modular, drop-in replacement.

%% file: fig-tex/eval_production_e2e_cdf.tex
\begin{figure*}[!htp]    
\centering
\includegraphics[width=\linewidth]{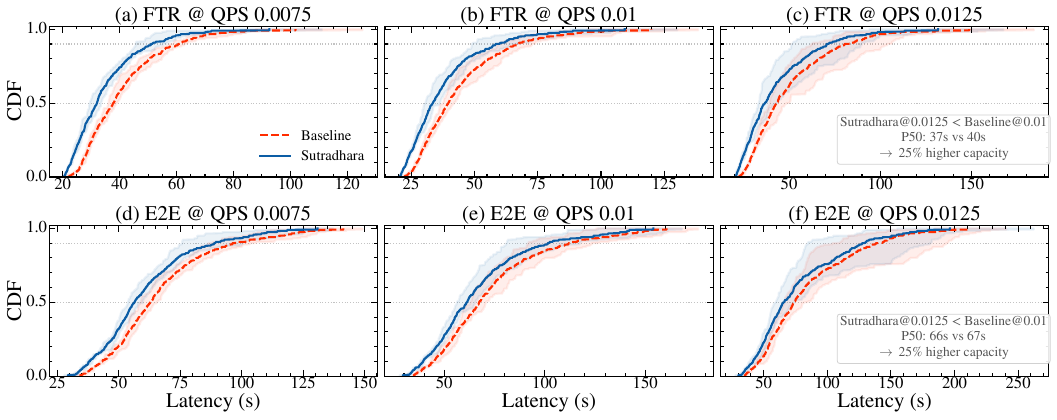}
\caption{Comparison of FTR and E2E Latency on the sampled Production trace}
    \label{fig:eval:production_trace}
\end{figure*}

%% file: fig-tex/gpt4_60_bars.tex

\begin{figure}[H]    
\centering
\includegraphics[width=\linewidth]{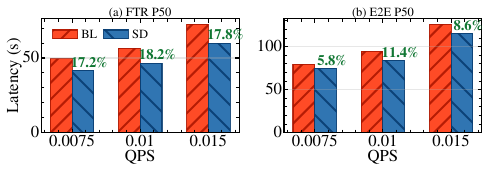}
\caption{Median FTR and E2E latency across QPS levels on a another subset of the production trace. \sysname consistently reduces FTR by 17--18\% and E2E by 6--11\% across all load levels.}
    \label{appendix:gpt4-bars}
\end{figure}